LA-UR-

Title:

Author(s):

Intended for:

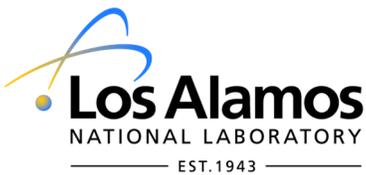

# $^{225}$Ac and $^{223}$Ra Production via 800 MeV Proton Irradiation of Natural Thorium Targets

J.W. Weidner*, S.G. Mashnik, K.D. John, B. Ballard, E.R. Birnbaum, L.J. Bitteker, A. Couture,

M.E. Fassbender, G.S. Goff, R. Gritzo, F.M. Hemez, W. Runde, J.L. Ullmann, L.E. Wolfsberg, and

F.M. Nortier[1]

*Los Alamos National Laboratory, P.O. Box 1663, Los Alamos, New Mexico, 87545, United States*

**Abstract**

Cross sections for the formation of $^{225,227}$Ac, $^{223,225}$Ra, and $^{227}$Th via the proton bombardment of natural thorium targets were measured at a nominal proton energy of 800 MeV. No earlier experimental cross section data for the production of $^{223,225}$Ra, $^{227}$Ac and $^{227}$Th by this method were found in the literature. A comparison of theoretical predictions with the experimental data shows agreement within a factor of two. Results indicate that accelerator-based production of $^{225}$Ac and $^{223}$Ra is a viable production method.

\* Corresponding author. Tel.: +1 505-665-6956.
 *E-mail addresses*: john.weidner@us.army.mil (J.W. Weidner), meiring@lanl.gov (F.M. Nortier).
[1] Principle investigator

# 1. Introduction

Alpha emitters have great advantages over beta emitters for use in cancer therapy due to a higher linear energy transfer and the limited alpha range in tissue. The increased availability and improved radiochemistry of alpha-particle-emitting nuclides for targeted therapy have presented new possibilities for their use in radioimmunotherapy. A number of alpha-particle-emitting nuclides, displaying half-lives ranging from minutes to days, are considered for application in targeted therapy (Couturier et al., 2005). An accelerator-based production method for two such radionuclides, $^{225}$Ac and $^{223}$Ra, is addressed in this work.

## 1.1 $^{225}$Ac

The actinide radioisotope $^{225}$Ac has a half-life of 10 days, emits four alpha particles in its decay chain, and has recently gained importance for application in future treatment of metastatic cancer via targeted alpha-immunotherapy (Miederer et al., 2008). $^{225}$Ac can also be used as a generator for $^{213}$Bi, another shorter-lived alpha emitter ($T_{1/2}$ = 45.6 min) considered for targeted alpha therapy.

Up till now, the widespread use of $^{225}$Ac and $^{213}$Bi in radiotherapy has been restricted by the limited availability of $^{225}$Ac. Presently, $^{225}$Ac is almost exclusively supplied by separating the isotope radiochemically from only two $^{229}$Th sources, one located at Oak Ridge National Laboratory (ORNL), USA (Boll et al., 2003) and the other at the Institute for Transuranium Elements in Karlsruhe (ITU), Germany (Apostolidis et al., 2001). The $^{229}$Th available at both sites was recovered from $^{233}$U, which has been in long-term storage at ORNL. This $^{233}$U was produced in kilogram quantities in the 1960's by neutron irradiation of $^{232}$Th in molten salt breeder reactors (Grimes, 1967).

## 1.2 $^{223}$Ra

A second radionuclide of interest, $^{223}$Ra, has a half-life of 11.4 days, also emits four alpha particles in its decay chain, and has demonstrated promise for the treatment of bone cancer (Bruland, et al., 2006). It can also be used as a generator for the production of $^{211}$Pb (Guseva, 2009).

Similar to $^{225}$Ac, the supply of $^{223}$Ra is limited due to its current method of production. Most commonly, $^{223}$Ra is milked from a $^{227}$Th (half-life = 18.7 days) cow, which in turn is obtained from the decay of the long-lived isotope $^{227}$Ac (half-life = 21.8 years). Only limited sources of $^{227}$Ac, collected from the decay of $^{235}$U or from the neutron irradiation of $^{226}$Ra (Kirby, et al., 2006), exist.

## 1.3 Alternate production methods

The expected growth in actinide isotope demand has led to the investigation of alternative production methods. For example, a reactor route for the production of $^{229}$Th, which alpha decays to $^{225}$Ra, relies upon multiple neutron captures on $^{226}$Ra (Van Geel, et al., 1994) and can be pursued at high flux reactors, such as the High Flux Isotope Reactor (HFIR) at ORNL. The evaluation of accelerator routes, including the production of $^{225}$Ac via the $^{226}$Ra (p,2n) and $^{226}$Ra (γ,n) nuclear reactions (Apostolidis, et al., 2005; Melville, et al., 2007), have also gained momentum.

This paper describes the investigation of another accelerator approach. This work is part of a wider evaluation of high-energy accelerator production routes employing intense 100 MeV, 200 MeV and 800 MeV proton beams and natural thorium targets for the large-scale production of the therapy isotopes $^{223,225}$Ra, $^{225,227}$Ac, and $^{227}$Th. Such beams are available at Los Alamos National Laboratory (LANL) and Brookhaven National Laboratory. A recent paper (Zhuikov, et al., 2011) describes a similar approach at energies below 200 MeV. Here we present cumulative cross section measurement results relevant to the production of these therapy isotopes with 800 MeV proton beams.

**2.     Experimental technique**

2.1     Irradiation

All aspects of this experiment were conducted at Los Alamos National Laboratory. Irradiation of the target foils occurred at the Los Alamos Neutron Science Center (LANSCE) accelerator facility. This accelerator is capable of delivering both H+ and H- ion beams at energies up to 800 MeV. This experiment utilized the H+ beam only. While the actual proton energy at the target foils was calculated to be 795.4 ± 0.5 MeV using stopping power data from Anderson and Ziegler (Anderson and Ziegler, 1977), the nominal value of 800 MeV is used throughout the text when referring to the proton energy.

Natural thorium foils of 99.7% purity were obtained from Goodfellow Corporation. Two foils were used as targets for this experiment, each measuring 2.5 cm x 2.5 cm and having thicknesses of 48.1 and 52.8 mg/cm$^2$. An aluminum beam monitoring foil with the same dimensions and a thickness of 65.3 mg/cm$^2$ was incorporated downstream of the thorium targets. The thickness of each foil was calculated by the ratio of its mass to surface area. Each foil was also measured with a micrometer at six different locations to ensure uniform thickness. All foils were sandwiched between single layers of Kapton tape with a thickness of 25 μm, which served as a catcher foil for recoil ions. After being mounted on acrylic frames, the target foils were irradiated together as a stack for approximately one hour without interruption at an average beam current of 67.6 nA. For beam monitoring purposes, the $^{27}$Al(p,x)$^{22}$Na reaction was applied. The cross section value for this reaction was obtained by examining published $^{27}$Al(p,x)$^{22}$Na cross section values at 800 MeV. Our search of the EXFOR database found eight values (Aleksandrov, et.al., 1996, Dittrich, et al., 1990, Heydegger, et al., 1976, Michel, et al., 1995, Morgan, et al., 2003, Taddeucci, et al., 1997, Titarenko, et al., 1999, Vonach, et al., 1997). A cross section value of 15.0 ± 0.9 mb was obtained by taking the average of those values within one standard deviation of the mean of all eight published values. The $^{27}$Al(p,x)$^{24}$Na reaction was also considered as a monitor reaction, but the predicted proton fluence showed a discrepancy of 15% when compared with that obtained from the $^{27}$Al(p,x)$^{22}$Na reaction. This discrepancy is believed to be due to secondary neutron contributions via the competing $^{27}$Al(n,α)$^{24}$Na reaction; hence, this monitor reaction was rejected.

2.2     Gamma and alpha spectroscopy

Several hours after end of bombardment (EOB), the foils were prepared for gamma spectroscopy. One thorium foil was initially counted on a PGT Ge(Li) detector with a relative efficiency of 13.7%. After 14 days, this foil was transferred to a LOAX-51370/20-S detector, which would enable the detection of the 40 keV gamma from $^{225}$Ra. The second foil was counted with an ORTEC GEM10P4-70 detector with a relative efficiency of 10%. All detectors used were well shielded and calibrated using NIST-tracable gamma calibration sources. The foils were counted repeatedly over a period of several months, and the decay curves of all isotopes were closely followed to ensure proper identification and to evaluate any possible interferences.

Analysis of the gamma spectra was conducted utilizing laboratory-specific analysis codes. The first of these codes, Raygun, is a Fortran-based analysis code developed by Ray Gunnink from Lawrence Livermore National Laboratory. GAMANAL is a more sophisticated analysis code that has been in use since the early 1970's (Gunnink and Niday, 1972). SPECANAL is a less rigorous version of GAMANAL that was coded at LANL in the 1980's in order to facilitate the batch processing of large numbers of gamma spectra. Gamma ray energies and intensities as listed on the Lund/LBNL Nuclear Data

Search website (Chu, et al., 1999) were used to determine the activity of the relevant isotopes (see Table 1).

Due to its very low activity, it was not possible to measure $^{227}$Ac nondestructively. Approximately 18 months after irradiation, the thorium foils were dissolved and the actinium isotopes chemically separated from the thorium matrix. Alpha counting samples were prepared by placing several stipples of a liquid fraction onto platinum discs and then evaporating the liquid to form a near-massless layer of alpha-emitting material. All alpha counting measurements were 10,000 minutes long and completed using ORTEC Octete+ eight-channel alpha spectrometers equipped with 300 mm$^2$ silicon detectors. These detectors were calibrated using NIST-traceable alpha calibration sources. Sample distance from the detector was maximized to achieve the best possible peak resolution, resulting in a counter efficiency of 4.9%.

Actinium and radium were separated from the thorium matrix using a modified process based on a technique developed at Oak Ridge National Laboratory (Boll, et al., 2005). The thorium foils were separated from their Kapton containment by soaking them in 25 mL of ethyl acetate. The ethyl acetate was decanted off and the two layers of Kapton were separated using tweezers. The ethyl acetate and Kapton were counted to ensure no material was removed from the foils. The thorium foil was then dissolved using 15M HNO$_3$ (40 mL) (Fisher Optima grade) and 1M HF (4 mL) (Fisher Optima grade). The foil was stirred in the HF/HNO$_3$ solution overnight. Following the dissolution process, a 5 µCi spike of $^{225}$Ac was added to the HF/HNO$_3$ mixture to measure the efficiency of the actinium chemical separation, and the thorium solution was taken to dryness using heat. The solution matrix was changed using three 50 mL washes of 8M HNO$_3$ whereby each wash was taken to dryness before the next was added. The residue from the matrix conversion steps was then dissolved in 2 mL of 8M HNO$_3$ and loaded onto a 10 mL bed volume BioRad MP1 anion exchange column. After loading the thorium solution, the actinium/radium fraction was collected with 30 mL of 8M HNO$_3$. This solution was taken to dryness, dissolved in 2 mL of 8M HNO$_3$ and passed through a second 10 mL bed volume anion column. The thorium was stripped from the first column with four 5 mL washes of 0.1M HNO$_3$ and collected for analysis.

Following the anion separation of actinium and radium from the bulk thorium matrix solution, a second 5 mL bed volume cation column was run following the above procedures. The actinium/radium fraction was taken to dryness and the matrix was converted to 0.1M HNO$_3$ by three 10 mL washes of 0.1M HNO$_3$, whereby each wash was taken to dryness before the next was added. The actinium was separated from radium by use of a BioRad AG 50WX4 cation column (10 mL bed volume pre-equilibrated with 0.1M HNO$_3$ prior to use). The radium/actinium residue was redissolved in 2 mL of 0.1M HNO$_3$ and loaded onto the column. The column was then washed four times with 5 mL of 0.1M HNO$_3$ and collected as the load fraction. Following the initial load and washing of the column, the radium was stripped from the column by four 5 mL washes of 1.2M HNO$_3$. Following the radium stripping, the actinium was eluted from the column by passing four 5 mL volumes of 8M HNO$_3$ and collecting the eluant. In order to ensure the complete separation of radium and actinium, the above process was repeated with the collected actinium fraction. Breakthrough of a small fraction of the thorium matrix into the final actinium fraction is observed, which results in interference from the daughter products of $^{228}$Th with the peaks of the immediate $^{227}$Ac daughters in the alpha spectrum. This complication was resolved by using the $^{215}$Po activity to quantify the $^{227}$Ac activity. As shown in Fig. 1, there is no interference with the 7,386 keV alpha peak associated with $^{215}$Po. Note that the $^{217}$At in Fig. 1 is from the decay of the $^{225}$Ac spike.

Since $^{227}$Ac decays to $^{215}$Po via the following path: $^{227}$Ac → $^{227}$Th → $^{223}$Ra → $^{219}$Rn → $^{215}$Po, there was concern that the alpha decay of $^{223}$Ra could provide enough energy to the recoiling $^{219}$Rn atoms to cause some of those gaseous atoms near the surface of the sample material to escape and drift into the vacuum system. Those atoms are assumed to be lost, leading to a measured $^{215}$Po activity that is artificially low. To prevent the loss of the gaseous $^{219}$Rn atoms, a thin collodion film was used to cover the sample during alpha counting. This technique has been successfully utilized by others to prevent recoil-ion contamination of surface-barrier detection systems while still enabling alpha counting (Inn, et al., 2009). The film is thin enough to allow the alpha particles to reach the detector with little degradation of energy resolution, but thick enough to retain the $^{219}$Rn atoms. A comparison of a count of the uncovered sample with an identical count of the same sample covered with a collodion film (Fig. 1) indicated that approximately 16% of the $^{219}$Rn atoms are lost from the uncovered sample.

2.3    Data analysis

It was recognized prior to this experiment that the non-destructive detection and quantification of some isotopes of interest via their gamma signatures alone would be a challenge. This anticipated complication, resulting mainly from the production of many fission products, was highlighted by theoretical predictions using the code MCNP6. It was anticipated that the X and gamma rays emitted by these fission products would likely interfere with the relatively weak gamma signatures of many of the isotopes of interest.

Initial results from the non-destructive counting of the thorium foils confirmed these predictions. Interference with the gamma signatures of $^{225}$Ra, $^{225}$Ac and $^{223}$Ra prevented their direct quantification. An example of the very busy spectrum below 125 keV is shown in Fig. 2. Because the 218.2 keV gamma emitted by $^{221}$Fr was without interference, the activities of $^{225}$Ac and $^{225}$Ra at EOB were quantified by evaluating very rigorously the parent-daughter in-growth and decay curve (Fig. 3) of $^{221}$Fr over time. A similar evaluation of the 351.06 keV gamma peak attributable to $^{211}$Bi enabled the quantification of $^{223}$Ra and $^{227}$Th. $^{227}$Th was also quantified directly by measuring its 235.971 keV gamma. The $^{227}$Th activities at EOB obtained from these two methods were within 5% of each other.

The $^{225}$Ac, $^{223,225}$Ra, and $^{227}$Th activities at EOB were calculated by performing a Markov chain Monte Carlo (MCMC) analysis (Higdon, et al., 2008; Hemez and Atamturktur, 2011) of the aforementioned decay curves. Using a Monte Carlo sampling technique, one million theoretical decay curves were fitted to each set of measured data points. An initial guess for the activity of each isotope at EOB was obtained from a least squares fit of its respective decay curve. Successive values for the radioisotope activities at EOB were obtained via a random walk technique in a range defined to be ± 30% of the initial guess values, and then used to generate successive decay curves. Each of these generated curves was assessed against how well it passed through the error bars of the data points associated with the measured decay curve for each isotope of interest, with all decay curves containing 10-20 measurements. Those curves that did not satisfactorily represent the measured data, as determined by preset variance values within the code, were rejected. Upon completion of one million satisfactory decay curves, the most probable value of the activity of each isotope at EOB, and the standard deviation of its probability distribution, were extracted. This process offers a statistically rigorous quantification of the uncertainties associated with the activity of each isotope at EOB. The 1 $\sigma$ relative uncertainties of the MCMC calculated activities at EOB ranged from 1.9% to 9.6%.

The uncertainty in each measured data point used in the MCMC analysis included the estimated uncertainty in detector calibration (2.1%), counting geometry (1%), branching ratio, and peak area (2-10%). These uncertainties were summed in quadrature to obtain a total uncertainty for each measured data point. The total uncertainty in each measured cross section was calculated by summing in quadrature the individual contributing uncertainties from the foil thickness (0.3% and 0.8%), integrated proton current (6.7%), decay constant, and radioisotope activities at EOB. Additionally, the total uncertainty in the $^{227}$Ac cross section value obtained from alpha counting includes assessments for the uncertainties introduced by the chemical separation process (4.7%). The uncertainty associated with each cross section value is expressed as a 1 $\sigma$ confidence level.

## 3. Results and Discussion

### 3.1 Cumulative Cross Sections

The proton irradiation of a $^{232}$Th foil will lead to the production of not only the various isotopes of interest, but of their relatively short-lived parents as well. For example, the decay of $^{229}$Pa and $^{225}$Th produced within the thorium targets will contribute to the measured activity of $^{225}$Ac; in the same way, the decay of $^{223}$Fr and $^{223}$Ac will contribute to the measured activity of $^{223}$Ra. Since the shorter-lived parents will contribute to the activity of all isotopes that we measured, the cross sections presented in this experiment are considered cumulative.

Theoretical cumulative cross section values for 800 MeV proton irradiation of thin $^{232}$Th targets were also calculated for the production of $^{223,225}$Ra, $^{225,227}$Ac and $^{227}$Th. These calculations utilized the MCNP6 code (Goorley et al, 2011) with three different event-generators: the Cascade-Exciton Model as implemented in the code CEM03.03 (Mashnik et al., 2008), the Bertini+MPM+Dresner+RAL event-generator (Bertini, 1969; Prael and Bozoian, 1988; Dresner, 1981; Atchison 1980), and the Liege intranuclear cascade model INCL (Boudard et al., 2002) merged with the ABLA evaporation/fission model (Junghans et al., 1998). Standard default values of all parameters were used in each model. The theoretical cross section values as calculated initially with MCNP6 were independent values and did not include contributions from any parent isotopes. Cumulative theoretical cross sections were derived thereafter, and include the appropriate contributions from the parent isotopes specified in sections 3.1.1 and 3.1.2 below.

With the exception of $^{225}$Ac, the data reported here represents the first measurement of these reactions at 800 MeV. The measured cumulative cross sections for the production of $^{225,227}$Ac, $^{223,225}$Ra, and $^{227}$Th via 800 MeV proton irradiation of thin, natural thorium foils are shown in Table 2. The table also compares the theoretical predictions with the measured values. Note that the predicted values show a factor of two agreement or better with experiment; however, no one model consistently yields a set of theoretical values that most closely matches all of the cross sections measured in this work. This prompted MCNP6 calculations using several event-generators and provides insights about the uncertainty of calculated results due to the physics ingredients emphasized by different models.

### 3.1.1 $^{232}$Th(p,x)$^{225}$Ac, $^{232}$Th(p,x)$^{225}$Ra and $^{232}$Th(p,x)$^{227}$Ac reactions

The $^{232}$Th(p,x)$^{225}$Ac cross sections account for the decay of $^{229}$Pa and $^{225}$Th adding to the measured activity of $^{225}$Ac. This measurement is well approximated by the three theoretical predictions. Conversely, the 20.3 ± 5.1 mb value obtained experimentally by Titarenko, et al. (2003) is well above our measurement.

The measured $^{232}$Th(p,x)$^{225}$Ra cross section in Table 2 is well forecasted by the Bertini model. The INCL and CEM models over predict and under predict, respectively, the measured value by roughly a factor of 2. The measured cross section for the impurity $^{227}$Ac, which includes a contribution from the decay of $^{227}$Ra, is 40.5% larger than the value measured for $^{225}$Ac, and is accurately predicted by the INCL model.

Chemical separation of the thorium foil proved vital to the quantification of $^{227}$Ac and the measurement of its cross section. This was not unexpected as a study of the gamma emissions of $^{227}$Ac and its daughter products clearly indicates that they are difficult isotopes to identify and quantify by gamma spectroscopy, especially in the presence of numerous fission products. An examination of the $^{227}$Ac progeny suggests that $^{211}$Pb might be measureable by reducing background through counting of the

427.088 keV (1.76% intensity) and the 404.853 keV (3.78% intensity) gammas in coincidence. Implementation of this counting method will be considered for future work.

### 3.1.2  $^{232}$Th(p,x)$^{223}$Ra and $^{232}$Th(p,x)$^{227}$Th reactions

The measured and theoretical cross sections in Table 2 for $^{232}$Th(p,x)$^{223}$Ra include contributions from the decay of $^{223}$Fr and $^{223}$Ac. The prediction from the Bertini model is in excellent agreement with the measurement. Similarly, the $^{232}$Th(p,x)$^{227}$Th cross sections include contributions from the decay of $^{229}$Pa and $^{225}$Th. In this case, only the median of the theoretical predictions is in very good agreement with our measurement, with the Bertini and INCL values approximately 50% less than our data and the CEM value approximately 50% greater.

## 3.2  Production rates and yields

Assuming an uninterrupted 10-day irradiation of a thorium target with thickness 5 g/cm$^2$ using a proton current of 1250 µA, the instantaneous production rate and total yield for each isotope is shown in Table 3. Such an irradiation scheme will be possible at the future Material Test Station being developed at LANSCE.

### 3.2.1  Production of $^{225}$Ac

This research demonstrates that intense proton accelerators can produce $^{225}$Ac in at least two distinct ways: directly via nuclear reactions and decay of short-lived parents, or indirectly via the decay of its longer-lived parent $^{225}$Ra. A third route is also possible by the path $^{229}$Pa (T$_½$ = 1.50 d) → $^{229}$Th (T$_½$ = 7,340 y) → $^{225}$Ra. Though this third production route was investigated, we were unable to quantify $^{229}$Pa and $^{229}$Th by either gamma or alpha spectroscopy due to interferences and very low activity, respectively.

The illustrated 10-day irradiation anticipates direct production of more than 20 Ci of $^{225}$Ac. Other undesirable isotopes of actinium would also be produced, but only one, $^{227}$Ac, has a half-life greater than $^{225}$Ac. From a $^{225}$Ac production perspective, the co-production of this long-lived $^{227}$Ac impurity is a concern; however, the expected $^{227}$Ac yield of 0.05 Ci represents only 0.25% of the total activity of the $^{225,227}$Ac combination. Additional research into the biological fate of the labeled carrier, the metabolized molecular products, and free actinium atoms is needed to ultimately determine if the presence of $^{227}$Ac would preclude accelerator produced $^{225}$Ac as a viable option for targeted alpha therapy.

Secondly, the several Curies of $^{225}$Ra yield from the production example could be utilized as a pure $^{225}$Ac generator. Although other isotopes of radium are also created in a production run, only two, $^{227,228}$Ra, beta-decay to actinium. The $^{227}$Ra has a relatively short 42.2 min half-life, but decays to the long-lived impurity $^{227}$Ac. Conversely, the relatively long-lived $^{228}$Ra has a half-life of 5.75 years, but decays to the short-lived isotope $^{228}$Ac (T$_½$ = 6.15 h), which in turn beta-decays to the alpha emitter $^{228}$Th (T$_½$ = 1.9 y). The impact of $^{227}$Ra production could be mitigated by delaying the chemical separation of radium until after the bulk of this isotope has decayed into $^{227}$Ac. Given its 5.75 y half-life, it is anticipated that the $^{228}$Ra would contribute comparatively little dose over the useful lifetime of a $^{225}$Ra generator.

### 3.2.2  Production of $^{223}$Ra

Three viable $^{223}$Ra production paths are evident from this work. Firstly, the direct production of $^{223}$Ra via nuclear reactions and decay of short-live parent nuclides is expected to yield more than a half-dozen Curies as shown in Table 3. Of the contaminant isotopes of radium that would be co-produced, only three are of concern: $^{225,226,228}$Ra. The 1600 y half-life of $^{226}$Ra may make its presence tolerable. The presence of $^{225,228}$Ra, however, may or may not be tolerable. Further research is needed to better understand the biologic consumption of such a radium-labeled carrier. In either case, accelerator-based production of $^{223}$Ra as a generator of $^{211}$Pb appears viable.

Secondly, the example irradiation scheme is expected to create nearly 11 Ci of $^{227}$Th. Though other alpha-emitting isotopes of thorium would also be produced, those thorium isotopes with mass numbers less than 227 all have half-lives shorter than 31 min, while the thorium isotopes with mass numbers larger than 227 all have half-lives greater than 1.9 y. Hence, proper timing of the chemical separation of thorium would lead to the recovery of a relatively high quality $^{227}$Th product that could serve as a multi-Curie generator for pure $^{223}$Ra.

The third $^{223}$Ra production route is from the creation of $^{227}$Ac, the long-lived parent of $^{227}$Th. Though the example production run predicts a 0.05 Ci yield of $^{227}$Ac, a longer irradiation of one year will produce 1.8 Ci. Since the half-life of $^{227}$Ac is nearly 22 years, such a $^{227}$Ac source would essentially serve as a permanent $^{227}$Th generator.

## 4. Summary and Conclusion

This work reports, for the first time, measured cross sections for the production of $^{223,225}$Ra, $^{227}$Th and $^{227}$Ac from the bombardment of a thorium target with 800 MeV protons. It also refines the uncertainty of the previously published $^{225}$Ac cross section to a more precise measurement. Extensive medical trials for $^{225}$Ac and $^{223}$Ra are currently constrained by the limited worldwide production of these radioisotopes. Based upon our measured cross section values and calculated yields, accelerator-based production of $^{223}$Ra and $^{225}$Ac with intense 800 MeV proton beams shows great promise. In fact, the predicted $^{225}$Ac yield from a single irradiation (20.2 Ci, Table 3) is nearly twenty times the current annual worldwide production. Furthermore, the expected 6.5 Ci yield of $^{223}$Ra would be a significant addition to the international inventory available for clinical trials and treatment. Curie amounts of the parent isotopes $^{225}$Ra and $^{227}$Th are also simultaneously produced, enabling the production of additional Curies of $^{225}$Ac and $^{223}$Ra from the resulting generators. Long-lived impurities such as $^{227}$Ac and $^{228}$Ra are also co-produced, however, and their presence may or may not be tolerable in therapy applications. The production of $^{227}$Ac may be advantageous, though, as a "permanent" generator of $^{227}$Th, $^{223}$Ra, and $^{211}$Pb.

Comparison of theoretical model predictions to our measured values shows reasonable-to-good agreement, with all of our measurements positioned near the median of the predicted values. The range of theoretical predictions is generally large, however, varying by factors of 2-5 for all isotopes except $^{225}$Ac. Additionally, no one model was shown to be consistently more accurate than another.

Given the promising results of this research, similar experiments are in progress at the LANCE facility to measure the production cross sections for these same isotopes using protons in the energy range below 200 MeV. The results of these experiments will be published in future articles.

**Acknowledgements**

We gratefully acknowledge funding by the United States Department of Energy Office of Science via an award from The Isotope Development and Production for Research and Applications subprogram in the Office of Nuclear Physics. We are also very thankful for the technical assistance provided by members of the LANL C-NR and C-IIAC groups, members of the AOT-OPS group, and the LANSCE-WNR staff.

# References

Aleksandrov, Y.V., Vasilev, S.K., Ivanov, R.B., Krizhanskij, L.M., Mikhaylova, M.A., Prikhodtseva, V.P., Eismont, V.G., 1996. (p, x) Reaction Cross Section on Aluminum at Intermediate Energy of Protons, Proc. 46th Ann. Conf. Nucl. Spectrosc. Struct. At. Nuclei, Moscow, p. 223.

Andersen, H.H., Ziegler, J.F., 1977. Hydrogen stopping powers and ranges in all elements. Pergamon, New York.

Apostolidis, C., Carlos-Marquez, R., Janssens, W., Molinet, R., Nikula, T., Ouadi, A., 2001. Cancer treatment using Bi-213 and Ac-225 in radioimmunotherapy. Nuclear News 44, 29-33.

Apostolidis, C., Molinet, R., Rasmussen, G., Morgenstern, A., 2005. Production of Ac-225 from Th-229 for targeted alpha therapy. Analytical Chemistry 77, 6288-6291.

Atchison, F., 1980. Spallation and fission in heavy metal nuclei under medium energy proton bombardment, in: Bauer, G. S. (Ed.), Proc. Meeting on targets for neutron beam spallation source, 1979, pp. 17-46, Conf-34, Kernforschungsanlage Julich GmbH, Julich, Germany.

Bertini, H.W., 1969. Intranuclear-cascade calculations of secondary nucleon spectra from nucleon-nucleus interactions in energy range 340 to 2900 MeV and comparison with experiment. Physical Review 188, 1711-1730.

Boll, R.A., Malkemus, D., Mirzadeh, S., 2005. Production of actinium-225 for alpha particle mediated radioimmunotherapy. Applied Radiation and Isotopes 62, 667-679.

Boll, R.A., Malkemus, D.W., Mirzadeh, S., 2003. Production of Ac-225 for alpha-particle-mediated radioimmunotherapy, 225th American Chemical Society National Meeting, New Orleans, USA.

Boudard, A., Cugnon, J., Leray, S., Volant, C., 2002. Intranuclear cascade model for a comprehensive description of spallation reaction data. Physical Review C 66, 044615.

Bruland, O.S., Nilsson, S., Fisher, D.R., Larsen, R.H., 2006. High-linear energy transfer irradiation targeted to skeletal metastases by the alpha-emitter Ra-223: Adjuvant or alternative to conventional modalities? Clinical Cancer Research 12, 6250S-6257S.

Chu, S.Y.F., Firestone, R.B., Eckstrom, L.P., 1999. WWW Table of Radioactive Isotopes, Version 2.0, URL: http:// nucleardata.nuclear.lu.se/nucleardata/toi/.

Couturier, O., Supiot, S., Degraef-Mougin, M., Faivre-Chauvet, A., Carlier, T., Chatal, J.F., Davodeau, F., Cherel, M., 2005. Cancer radioimmunotherapy with alpha-emitting nuclides. European Journal of Nuclear Medicine and Molecular Imaging 32, 601-614.

Dittrich, B., Herpers, U., Lupke, M., Michel, R., Hofmann, H.J., Wolfli, W., 1990. Determination of cross-sections for the production of Be-7, Be-10, and Na-22 by high-energy protons. Radiochimica Acta 50, 11-18.

Dresner, L., 1981. EVAP-A Fortran Program for Calculating the Evaporation of Various Particles from Excited Compound Nuclei, ORNL-TM-7882, Oak Ridge, USA.

Goorley, T., James, M., Booth, T., Brown, F., Bull, J., Cox, L. J., Durkee, J., Elson, J., Fensin, M., Forster, R. A., Hendricks, J., Hughes, H. G., Johns, R. B., Kiedrowski, R. Martz, Mashnik, S., McKinney, G., Pelowitz, D., Prael, R., Sweezy, J., Waters, L., Wilcox, T., Zukaitis, T., 2011. Initial MCNP6 release overview, MCNP6 version 0.1. LA-UR-11-05198, to be published in the journal Nuclear Technology.

Grimes, W.R., 1967. Chemical Research and Development for Molten-Salt Breeder Reactors. Report, ORNL-TM-1853, Oak Ridge, Tennessee, USA.

Gunnink, R., Niday, J. B., 1972. Computerized Quantitative Analysis by Gamma-ray Spectrometry. Vol I. UCRL-51061, Livermore, California, USA.

Guseva, L.I., 2009. A tandem generator system for production of (223)Ra and (211)Pb/(211)Bi in DTPA solutions suitable for potential application in radiotherapy. Journal of Radioanalytical and Nuclear Chemistry 281, 577-583.

Hemez, F.M., Atamturktur, S., 2011. The dangers of sparse sampling for the quantification of margin and uncertainty. Reliability Engineering & System Safety 96, 1220-1231.

Heydegger, H.R., Turkevich, A.L., Vanginneken, A., Walpole, P.H., 1976. Production of Be-7, Na-22, and Mg-28 from Mg, Al, and SiO2 by protons between 82 and 800 MeV. Physical Review C 14, 1506-1514.

Higdon, D., Gattiker, J., Williams, B., Rightley, M., 2008. Computer model calibration using high-dimensional output. Journal of the American Statistical Association 103, 570-583.

Inn, K.G.W., Hall, E., Woodward, J.T., Stewart, B., Pollanen, R., Selvig, L., Turner, S., Outola, I., Nour, S., Kurosaki, H., LaRosa, J., Schultz, M., Lin, Z., Yu, Z., McMahon, C., 2008. Use of thin collodion films to prevent recoil-ion contamination of alpha-spectrometry detectors. Journal of Radioanalytical and Nuclear Chemistry 276, 385-390.

Junghans, A.R., de Jong, M., Clerc, H.G., Ignatyuk, A.V., Kudyaev, G.A., Schmidt, K.H., 1998. Projectile-fragment yields as a probe for the collective enhancement in the nuclear level density. Nuclear Physics A 629, 635-655.

Kirby, H.W., Morss, L.R., 2006. Actinium, in: Morss, L.R., Edelstein, N.M., Fuger, J. (Eds.), The chemistry of the actinide and transactinide elements, 3 ed. Springer, The Netherlands, pp. 18-24.

Mashnik, S. G., Gudima, K. K., Prael, R. E., Sierk, A. J., Baznat, M. I., Mokhov, N. V., 2008. CEM03.03 and LAQGSM03.03 Event Generators for the MCNP6, MCNPX, and MARS15 Transport Codes. Invited lectures presented at the Joint ICTP-IAEA Advanced Workshop on Model Codes for Spallation Reactions, ICTP, Trieste, Italy, LA-UR-08-2931, Los Alamos; E-print: arXiv:0805.0751.

Melville, G., Meriarty, H., Metcalfe, P., Knittel, T., Allen, B.J., 2007. Production of Ac-225 for cancer therapy by photon-induced transmutation of Ra-226. Applied Radiation and Isotopes 65, 1014-1022.

Michel, R., Gloris, M., Lange, H.J., Leya, I., Lüpke, M., Herpers, U., Dittrich-Hannen, B., Rösel, R., Schiekel, T., Filges, D., Dragovitsch, P., Suter, M., Hofmann, H.J., Wölfli, W., Kubik, P.W., Baur, H., Wieler, R., 1995. Nuclide production by proton-induced reactions on elements ($6 \leq Z \leq 29$) in the energy range from 800 to 2600 MeV. Nuclear Instruments and Methods in Physics Research Section B: Beam Interactions with Materials and Atoms 103, 183-222.

Miederer, M., Scheinberg, D.A., McDevitt, M.R., 2008. Realizing the potential of the Actinium-225 radionuclide generator in targeted alpha particle therapy applications. Advanced Drug Delivery Reviews 60, 1371-1382.

Morgan, G.L., Alrick, K.R., Saunders, A., Cverna, F.C., King, N.S.P., Merrill, F.E., Waters, L.S., Hanson, A.L., Greene, G.A., Liljestrand, R.P., Thompson, R.T., Henry, E.A., 2003. Total cross sections for the production of Na-22 and Na-24 in proton-induced reactions on Al-27 from 0.40 to 22.4 GeV. Nuclear Instruments & Methods in Physics Research Section B-Beam Interactions with Materials and Atoms 211, 297-304.

Prael, R.E., Bozoian, M., 1988. Adaptation of the Multistage Preequilibrium Model for the Monte Carlo Method. LANL report LA-UR-88-3238, Los Alamos, USA.

Taddeucci, T.N., Ullmann, J., Rybarcyk, L.J., Butler, G.W., Ward, T.E., 1997. Total cross sections for production of Be-7, Na-22, and Na-24 in p+Li-7 and p+Al-27 reactions at 495 and 795 MeV. Physical Review C 55, 1551-1554.

Titarenko, Yu. E., Shvedov, O.V., Batyaev, V.F., Karpikhin, E.I., Zhivun, V.M., Mulambetov, R.D., Sosnin, A.N., Mashnik, S.G., Prael, R.E., Gabriel, T.A., Blann, M., 1999. Experimental and Computer Simulation Study of Radionuclide Production in Heavy Materials Irradiated by Intermediate Energy Protons. LANL Report LA-UR-99-4489, Los Alamos (1999); Proc. Embedded 3rd Topical Meeting on Nuclear Applications of Accelerator Technology (AccApp'99), ANS 1999 Winter Meeting, Long Beach, CA, November 14-18, 1999, ANS (1999) pp. 212-221; E-print: nucl-ex/9908012.

Titarenko, Yu.E., Batyaev, V.F., Karpikhin, E.I., Mulambetov, R.D., Koldobsky, A.B., Zhivun, V.M., Mulambetova, S.V., Lipatov, K.A., Nekrasov, Yu.A., Belkin, A.V., Alexeev, N.N., Schegolev, V.A., Goryachev, Yu.M., Luk'yashin, V.E., Firsov, E.N., 2003. Experimental and theoretical study of the yields of residual product nuclei produced in thin targets irradiated by 100-2600 MeV protons, INDC(CCP)-434, pp. 73-79.

Van Geel, J.N.C., Fuger, J.J., Koch, L., 1994. Method for producing actinium-225 and bismuth-213. United States Patent 5355394.

Vonach, H., Pavlik, A., Wallner, A., Drosg, M., Haight, R.C., Drake, D.M., Chiba, S., 1997. Spallation reactions in Al-27 and Fe-56 induced by 800 MeV protons. Physical Review C 55, 2458-2467.

Zhuikov, B.L., Kalmykov, S.N., Ermolaev, S.V., Aliev, R.A., Kokhanyuk, V.M., Matushko, V.L., Tananaevc, I.G., Myasoedov, B.F., 2011. Production of $^{225}$Ac and $^{223}$Ra by irradiation of Th with accelerated protons. Radiokhimiya 53, 73-80.

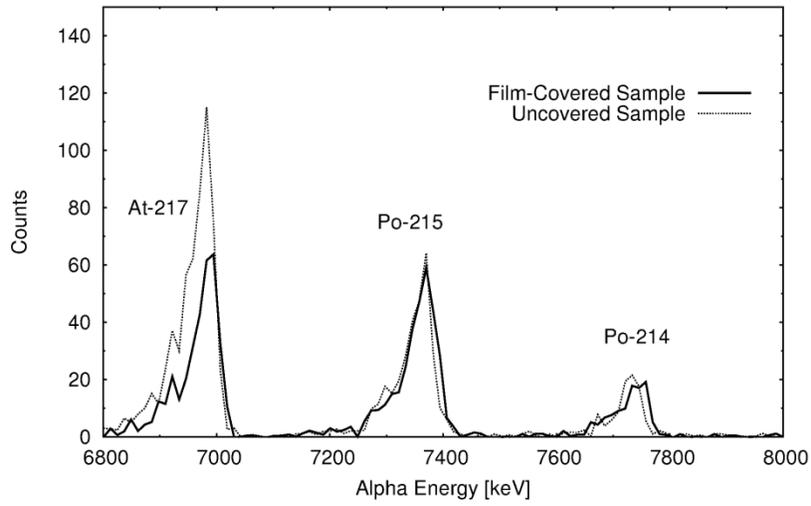

**FIG. 1.** Comparison of data from two identical counts of the same sample, one covered by a thin collodion film and one uncovered. The activity of $^{215}$Po was used to quantify the activity of $^{227}$Ac at end of bombardment.

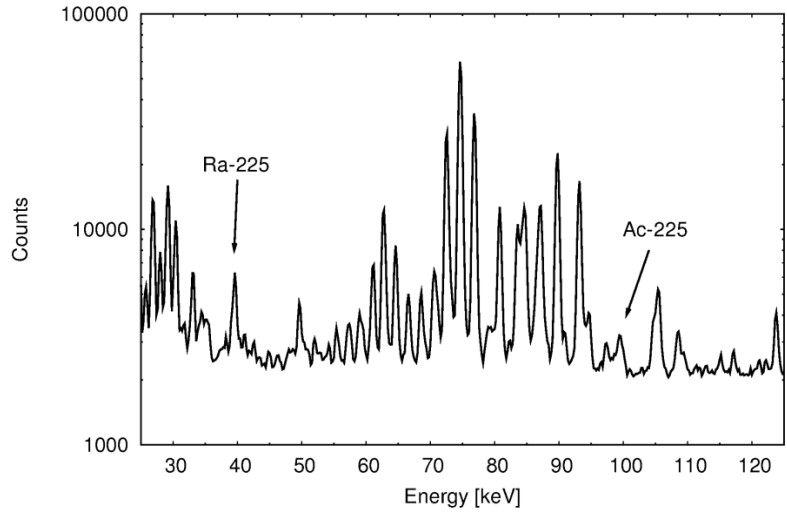

**FIG. 2.** Example of the significant X- and gamma ray interference below 125 keV.

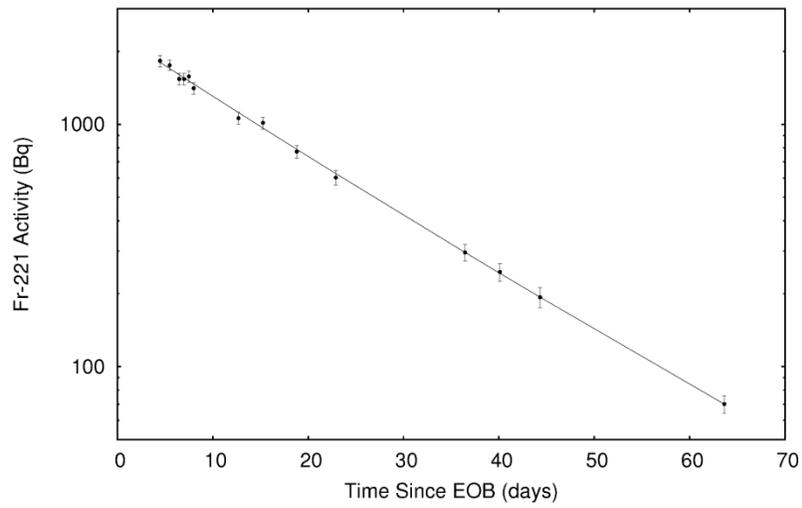

**FIG. 3.** Measurements of the $^{221}$Fr activity made over many weeks. The solid line represents the activity of $^{221}$Fr based upon the $^{225}$Ac and $^{225}$Ra activities at EOB as determine by a Markov chain Monte Carlo analysis.

**Table 1**
Nuclear decay data utilized for the analysis of radioisotopes relevant to this work (Chu, et al., 1999).

|  | Half-Life | Decay Mode (%) | $E_\gamma$ [keV] | $I_\gamma$ [%] | $E_\alpha$ [keV] | $I_\alpha$ [%] |
|---|---|---|---|---|---|---|
| $^{225}$Ac | 10.0 d | α (100) | 99.91 | 1.01 | 5,830 | 50.7 |
| $^{225}$Ra | 14.9 d | β⁻ (100) | 40.09 | 30 | - | - |
| $^{227}$Ac | 21.773 y | β⁻ (98.620) | 100 | 0.009 | 4,953.26 | 47.7 |
|  |  | α (1.380) |  |  | 4,940.7 | 39.6 |
| $^{223}$Ra | 11.435 d | α (100) | 269.459 | 13.7 | 5,716.23 | 52.6 |
|  |  |  |  |  | 5,606.73 | 25.7 |
| $^{227}$Th | 18.72 d | α (100) | 235.971 | 12.3 | 6,038.01 | 24.2 |
|  |  |  | 256.25 | 7.0 | 5,977.72 | 23.5 |
| $^{221}$Fr | 4.9 m | α (100) | 218.19 | 11.6 | 6,341 | 83.4 |
| $^{211}$Bi | 2.14 m | α (99.724) | 351.059 | 12.91 | 6,622.9 | 83.77 |
| $^{219}$Rn | 3.96 s | α (100) | 271.23 | 10.8 | 6,819.1 | 79.4 |
| $^{215}$Po | 1.781 ms | α (99.99977) | 438.8 | 0.04 | 7,386.2 | 100 |

**Table 2**
Comparison of theoretical predictions to experimentally measured cumulative cross sections for the formation of $^{223,225}$Ra, $^{225,227}$Ac and $^{227}$Th.

|  | Measured Value [mb] | Bertini Value [mb] | CEM Value [mb] | INCL Value [mb] |
|---|---|---|---|---|
| $^{225}$Ac | 14.8 ± 1.1 | 12.5 | 15.0 | 12.1 |
| $^{225}$Ra | 3.3 ± 0.4 | 3.8 | 1.5 | 7.8 |
| $^{227}$Ac | 20.8 ± 2.0 | 14.5 | 11.0 | 21.5 |
| $^{223}$Ra | 5.3 ± 0.6 | 5.3 | 3.1 | 9.6 |
| $^{227}$Th | 12.7 ± 0.7 | 6.3 | 18.6 | 7.4 |

**Table 3**
Production rates and projected yields from a 10-day irradiation of a 5 g/cm² natural thorium target at the future Los Alamos National Laboratory Material Testing Station using a nominal 800 MeV proton beam at 1250 µA.

|  | Production Rate[a] [µCi/µA·h] | Yield [Ci] |
|---|---|---|
| $^{225}$Ac | 93.6 | 20.2 |
| $^{225}$Ra | 14.0 | 3.4 |
| $^{227}$Ac | 0.17 | 0.05 |
| $^{223}$Ra | 29.3 | 6.6 |
| $^{227}$Th | 42.9 | 10.8 |

[a] Instantaneous production rate, which does not account for decay